# Anisotropic interactions of a single spin and dark-spin spectroscopy in diamond

R. J. Epstein, F. M. Mendoza, Y. K. Kato & D. D. Awschalom*

Center for Spintronics and Quantum Computation, University of California, Santa Barbara, California 93106, USA
*e-mail: awsch@physics.ucsb.edu

**The nitrogen-vacancy (N-V) center in diamond is a promising atomic-scale system for solid-state quantum information processing. Its spin-dependent photoluminescence has enabled sensitive measurements on *single* N-V centers, such as: electron spin resonance[1], Rabi oscillations[2], single-shot spin readout[3] and two-qubit operations with a nearby $^{13}$C nuclear spin[4]. Furthermore, room temperature spin coherence times as long as 58 μs have been reported for N-V center ensembles[5]. Here, we have developed an angle-resolved magneto-photoluminescence microscopy apparatus to investigate the anisotropic electron spin interactions of single N-V centers at room temperature. We observe negative peaks in the photoluminescence as a function of both magnetic field magnitude and angle that are explained by coherent spin precession and anisotropic relaxation at spin level anti-crossings. In addition, precise field alignment unmasks the resonant coupling to neighboring "dark" nitrogen spins that are not otherwise detected by photoluminescence. The latter results demonstrate a means of investigating small numbers of dark spins via a single bright spin under ambient conditions.**

The nitrogen-vacancy defect pair has an anisotropic electron spin Hamiltonian due to its trigonal symmetry[6] and spin-orbit coupling[7]. Consequently, the degree of spin level mixing and coupling to nearby impurity spins is very sensitive to the orientation of an applied magnetic field, which has not been controlled in previous experiments on single N-V centers. Figure 1a depicts the atomic structure and relevant energy levels of the (negatively charged) N-V center. The triplet ($^3$A) ground state[8,9,10] has a zero-field splitting between the $\left|m_S = 0\right\rangle$ and $\left|m_S = \pm 1\right\rangle$ sublevels quantized along the N-V symmetry axis, a $\langle 111 \rangle$ crystal axis[11]. Zero-field spin splittings have also been measured in the triplet ($^3$E) excited state but there is no consensus on the arrangement of sublevels[12,13,14,15]. Linearly polarized optical excitation of the $^3$A → $^3$E transition preferentially pumps the spin system into the $\left|0\right\rangle$ ground state sublevel[16]. In addition, the average photon emission rate is substantially smaller for transitions involving the $\left|\pm 1\right\rangle$ levels than for the $\left|0\right\rangle$ level[3], which enables the spin state to be determined by the photoluminescence (PL) intensity. Both of the latter two effects have been attributed to spin-dependent intersystem crossing to the singlet ($^1$A) level[17].

Here, the phonon-broadened transition between the $^3$E and $^3$A states is detected via non-resonant PL (see Methods). For example, Fig. 1b is a spatial image of the spectrally-integrated PL from a diamond sample with the laser focused roughly 1 μm below the surface. Multiple resolution-limited features are observed within the 20×20 μm field. In order to determine that a given feature is due to a single emitter, a histogram is plotted of the time τ between consecutive photon detection events using a Hanbury Brown and Twiss detection geometry[18], yielding the experimental intensity correlation function $g^{(2)}(\tau)$. Figure 1c shows data from a color center labeled NV1. The value of $g^{(2)}(0)$ is below 0.5, proving that NV1 is a single color center[19]. In addition, the rise of $g^{(2)}(\tau)$ above unity with increasing $|\tau|$ is indicative of intersystem crossing to the $^1A$ meta-stable state[19,20].

These single photon emitters are further characterized by optically detected electron spin resonance (ESR) (see Methods). Figure 1d contains data from NV1 with zero applied magnetic field. In this case, the $|\pm 1\rangle$ ground state levels are degenerate so that only one resonance ($m_S = 0 \rightarrow \pm 1$) is observed at 2.87 GHz, the characteristic zero-field splitting of N-V centers[11]. The data is fit with a Lorentzian of 11 MHz FWHM (solid line), which is comparable to previous measurements on similar diamond samples[1].

Previous strain-dependent optical measurements on ensembles of N-V centers indicated that electric dipole transitions are allowed for dipoles in the plane perpendicular to the symmetry axis[6]. Figure 2a depicts an N-V center with transition dipoles $\mathbf{X} \| [\bar{1}\,\bar{1}\,2]$ and $\mathbf{Y} \| [1\,\bar{1}\,0]$ in a (111) plane. The excitation is along [001] for all measurements, therefore, $I_{\mathrm{PL}}$ depends on the laser polarization angle $\phi$ due to unequal excitation of the two dipoles. The dependence of $I_{\mathrm{PL}}$ on $\phi$ is measured for twenty N-V centers and exhibits either vertical or horizontal lobes, as exemplified for three N-V centers in Fig. 2b. The measured anisotropies, defined as the ratio $I_{\mathrm{PL}}(\phi = 0°)/I_{\mathrm{PL}}(\phi = 90°)$, are roughly 2:1 and 1:2 on average; a simple anisotropy calculation gives 3:1, 3:1, 1:3, and 1:3 for the N-V symmetry axis along [111], $[\bar{1}\,\bar{1}\,1]$, $[1\,\bar{1}\,1]$ and $[\bar{1}\,1\,1]$, respectively (see Methods). The measured data are modified by the presence of polarized background PL (subtracted from the data) and a small dip, most visible for NV2 at $\phi = 0°$, which varies in depth from center to center. Nevertheless, comparison of the measured and calculated anisotropies enables the number of possible orientations of a given center to be reduced from four to two.

The double degeneracy of the polarization anisotropy is lifted by application of a magnetic field **B**. The upper panel of Fig. 2c shows the ground state spin levels as a function of |**B**| calculated using the Hamiltonian[7] $H = g\mu_B \mathbf{B}\cdot\mathbf{S} + D(S_z^2 - S(S+1)/3)$, where $\mu_B$ is the Bohr magneton, $g = 2.0$, $S = 1$ and $D = 2.88$ GHz. A level anti-crossing (LAC) near 1000 G is predicted for an N-V center with symmetry axis at a small angle to **B** (6°; solid lines), but not for a large angle (55°; dashed lines). The middle panel is a plot of $|\alpha|^2$, the coefficient of $|0\rangle_z$, calculated for both angles and for each spin level, $|m_S\rangle = \alpha|0\rangle_z + \beta|-1\rangle_z + \gamma|+1\rangle_z$, where the z subscript denotes the [111] basis; the plot shows how the spin states mix with |**B**| and will be used to model the data below. The lower panel is a plot of $I_{PL}$ as a function of |**B**|, with a ~1° angle between **B** and [111], for the same three N-V centers as in Fig. 2b. Whereas NV1 and NV2 have similar polarization dependences, their field dependences reveal that the symmetry axis of NV1 (NV2) is (non-)parallel to **B**. For NV1, the negative peak at ~1000 G coincides with the calculated LAC. Additionally, NV1 exhibits a peak at ~500 G, the origin of which is less clear.

In order to investigate these peaks further, $I_{PL}$ is measured as a function of |**B**| with **B** at a series of angles θ in the $(1\bar{1}0)$ plane; a selection of such data from NV1 is displayed in Fig. 3a. As **B** approaches the [111] direction (θ → 0°), the LAC peak gets narrower, as expected from previous ensemble measurements[21,22]. At sufficiently small angles, however, the LAC peak amplitude decreases for NV1 (Fig. 3b) or even vanishes

for some centers, such as NV4 (Fig. 3c). This is in contrast to the ensemble measurements[21,22] that showed a maximum in the LAC peak amplitude at θ = 0°. According to the Hamiltonian above, however, spin mixing should vanish at θ = 0°. The presence of residual spin mixing at θ = 0° has been attributed to strain and nuclear interactions[23], which vary from center to center.

The LAC can be modeled for small values of θ by considering $|0\rangle$ and $|-1\rangle$ as pseudo-spin 1/2 states. With preferential population of $|0\rangle_z$, the $|+1\rangle$ level is ignored because it has a negligible overlap with $|0\rangle_z$ (see Fig. 2c, middle panel). We then have an effective Hamiltonian $H = g\mu_B(\mathbf{B}-\mathbf{B_0})\cdot\mathbf{s}$, where $\mathbf{B_0}$ accounts for the zero-field spin splitting and **s** is the pseudo-spin operator. The Bloch equations for **B** in the $(1\bar{1}0)$ plane are taken to be

$$\frac{ds_x}{dt} = -\Omega_z s_y - \frac{s_x}{T_2}$$
$$\frac{ds_y}{dt} = \Omega_z s_x - \Omega_x s_z - \frac{s_y}{T_2}$$
$$\frac{ds_z}{dt} = \Omega_x s_y - \frac{s_z}{T_1} + \Gamma$$

where $\hbar\Omega = g\mu_B(\mathbf{B}-\mathbf{B_0})$, $x\|[\bar{1}\,\bar{1}\,2]$, $y\|[1\,\bar{1}\,0]$, $z\|[111]\|\mathbf{B_0}$, Γ is the rate of optical spin orientation along $z$, and $T_1$ and $T_2$ are effective spin relaxation times that depend on Γ. In this model, the spin relaxation is anisotropic in that $T_1$ and $T_2$ are fixed relative to the crystal axes rather than the magnetic field. The steady-state solution for $s_z$ is

$$s_z = \frac{T_1\Gamma\left(1+\Omega_z^{\,2}T_2^{\,2}\right)}{1+\Omega_x^{\,2}T_1T_2+\Omega_z^{\,2}T_2^{\,2}}\ .$$

We then take $I_{PL} = An_0 + Bn_{-1} = A(1/2+s_z)+B(1/2-s_z)$, where $A$ and $B$ ($n_0$ and $n_{-1}$) are the PL rates (occupation probabilities) for the $\left|0\right\rangle_z$ and $\left|-1\right\rangle_z$ levels, respectively. This simple model describes the experimental data for a wide range of angles and magnetic fields, where the same fit parameters are used for all angles (Fig. 3b and c lines). The fits yield $T_1$ = 64 ns and $T_2$ = 11 ns for NV4 using a laser power of 2.9 mW (Fig. 3c) and $T_1$ = 130 ns and $T_2$ = 23 ns for a power of 890 μW (not shown). The results indicate that the laser introduces substantial anisotropic spin relaxation via excitation out of the ground state manifold.

The 500G peak evolves similarly to the LAC peak but over a broader angular range. It has nearly the same amplitude and width at a given field angle for all N-V centers investigated with suitable orientation. In addition, the peak has been observed in all four samples measured. The evolution of the 500G peak with θ and **B** can be reasonably accounted for in the model by postulating an LAC in the excited state, as is expected to occur at some field due to the presence of zero-field spin splittings[12,13,14]. Figure 3d shows the normalized amplitudes A of the 500G and LAC (1000 G) peaks as a function of θ. Fits of the model to this data (red lines) and to the field scans (not shown) yield $T_1$ = 36 ns and $T_2$ = 1.8 ns for an excited state LAC at ~500 G.

Similar data is taken on ensembles of N-V centers in Fig. 4a, where the 500G peak is found to reduce in amplitude by ~50% at θ = 0°. Additionally, higher resolution field scans around 500 G (Fig. 4a, inset) reveal the characteristic hyperfine quintuplet of substitutional nitrogen ($N_S$) centers[24]. These peaks appear when the electron spin splittings of $N_S$ and N-V centers are equal, resulting in enhanced cross relaxation via the magnetic dipole interaction[25,26,27]. Notably, the $N_S$ peak amplitudes diminish as |θ|

increases, suggesting that the 500G peak is commensurate with a decrease in spin polarization, like the LAC peak.

The above results indicate the most suitable angular range ($|\theta| < 1°$) for observing the coupling of a single N-V center to its neighboring $N_S$ spins. Figure 4b displays two field scans for NV4 at $\theta = 0°$ (blue circles) and 1° (red circles), showing the increase in peak magnitude with decreasing angle, as in the ensemble data. Figure 4c shows similar data for NV1 at $\theta = 0°$ with powers of 300 μW (blue circles) and 890 μW (red circles). The nitrogen lines broaden and decrease in relative amplitude at higher power, which may result from ionization of the $N_S$ centers[28]. The difference in the depth of the $N_S$ peaks between NV4 and NV1 is suggestive of variation in the local nitrogen concentration.

Finally, it is noteworthy that the $N_S$ centers are "dark" in that they are not directly detected by photoluminescence. By measuring a single N-V center, the number of $N_S$ spins that can be probed is decreased by orders of magnitude relative to ensemble measurements. Furthermore, this dark-spin spectroscopy technique is in principle applicable to a variety of paramagnetic defects in diamond. With higher purity samples[29] and single ion implantation[30], these results could make possible the long-range coupling of two individually addressable N-V centers connected by a chain of dark spins, enabling experimental tests of spin lattice theories and quantum information processing schemes.

## Acknowledgements

We thank O. Gywat for valuable discussions and G. C. Farlow for high energy electron irradiation of several samples. This work was supported by AFOSR, DARPA/MARCO, and ARO.

## Methods

### Sample preparation and characterization

The samples are commercially available high-temperature high-pressure diamond (Sumitomo) with nominal dimensions of 1.4×1.4×1.0 mm and nitrogen concentrations of 100-1000 ppm, measured by UV absorption[31]. N-V centers naturally exist in these samples with estimated densities ranging from $10^{10}$ to $10^{13}$ $cm^{-3}$. Two parallel sample faces are polished and have (100) orientation. For ensemble measurements, samples are irradiated with 1.7 MeV electrons with a dose of $5\times10^{17}$ $cm^{-3}$ and subsequently annealed at 900°C for 2 hours to increase the N-V center concentration[6].

### Experimental techniques

The measurement apparatus is based around a confocal microscope with a Hanbury Brown and Twiss detection scheme[18]. A diode-pumped solid state laser emitting at 532 nm is linearly polarized and focused onto the sample with a microscope objective of numerical aperture 0.73 and working distance 4.7 mm. The linear polarization is set to any desired angle by changing the retardance of a variable wave plate (fast axis at 45° to the initially vertical polarization) in front of a quarter-wave plate (fast axis vertical). The laser spot is positioned on the sample in both lateral dimensions with a fast steering mirror. The PL from the sample is collected by the same microscope objective, passed through a dichroic mirror and a 640 nm long-pass filter and sent via a 50/50 beam-splitter to two fiber-coupled silicon avalanche photodiode modules. For antibunching measurements, the outputs of the detectors are connected to a time-correlated single photon counting module. For measurements of N-V center ensembles, PL is detected with a photodiode and lock-in amplifier. In general, the PL is used as a feedback signal to compensate for thermal drift, enabling a single N-V center to be tracked for several days. The samples are at room temperature for all measurements discussed in this Letter.

The static magnetic field **B** is applied to the sample with a permanent magnet mounted on a multi-axis stage that allows the distance between the magnet and the sample to be adjusted by a stepper motor, thereby setting |**B**| at the sample. In addition, the polar and azimuthal angles of **B** are manually adjustable with micrometers while keeping |**B**| constant at the sample to better than 1%. For ESR measurements, a 25 μm diameter gold wire is connected to a microwave signal generator (with 16 dBm power output) and placed in close proximity (~50 μm) to the laser spot.

**Polarization anisotropy calculation**

According to Fermi's golden rule in the electric dipole approximation, the absorption rate is proportional to $|\mathbf{D}\cdot\mathbf{E}|^2$, where **D** is the dipole matrix element and **E** is the excitation electric field vector. The absorption anisotropy is estimated as $|\mathbf{Y}\cdot\mathbf{V}|^2/|\mathbf{X}\cdot\mathbf{H}|^2$, where **X** and **Y** are the dipoles (see Fig. 2a) and $\mathbf{V}\|\left[1\,\bar{1}\,0\right]$ and $\mathbf{H}\|[110]$ are polarization vectors. A more accurate calculation would account for the dipole radiation pattern, the large numerical aperture of the microscope objective and refractive effects

due to the diamond air boundary. However, the simple equation above is sufficient to account for both the anisotropy orientations and degeneracies that are measured.

## References

[1] Gruber, A. *et al.* Scanning confocal optical microscopy and magnetic resonance on single defect centers. *Science* **276**, 2012 (1997).

[2] Jelezko, F., Gaebel, T., Popa, I., Gruber, A. & Wrachtrup, J. Observation of coherent oscillations in a single electron spin. *Phys. Rev. Lett.* **92**, 076401 (2004).

[3] Jelezko, F. *et al.* Single spin states in a defect center resolved by optical spectroscopy. *Appl. Phys. Lett.* **81**, 2160 (2002).

[4] Jelezko, F. *et al.* Observation of coherent oscillation of a single nuclear spin and realization of a two-qubit conditional quantum gate. *Phys. Rev. Lett.* **93**, 130501 (2004).

[5] Kennedy, T. A., Colton, J. S., Butler, J. E., Linares, R. C. & Doering, P. J. Long coherence times at 300 K for nitrogen-vacancy center spins in diamond grown by chemical vapor deposition. *Appl. Phys. Lett.* **83**, 4190 (2003).

[6] Davies, G. & Hamer, M. F. Optical studies of the 1.945 eV vibronic band in diamond. *Proc. Roy. Soc. A* **348**, 285 (1976).

[7] Pryce, M. H. L. A modified perturbation procedure for a problem in paramagnetism. *Proc. Phys. Soc. A* **63**, 25 (1950).

[8] Reddy, N. R. S., Manson, N. B., & Krausz, E. R. Two-laser spectral hole burning in a colour centre in diamond. *J. Lumin.* **38,** 46 (1987).

[9] van Oort, E., Manson, N. B. & Glasbeek, M. Optically detected spin coherence of the diamond N-V centre in its triplet ground state. *J. Phys. C* **21,** 4385 (1988).

[10] Redman, D. A., Brown, S., Sands, R. H. & Rand, S. C. Spin dynamics and electronic states of N-V centers in diamond by EPR and four-wave-mixing spectroscopy. *Phys. Rev. Lett.* **67,** 3420 (1991).

[11] Loubser, J. H. N. & van Wyk, J. A. Electron spin resonance in the study of diamond. *Rep. Prog. Phys.* **41,** 1201 (1978).

[12] Redman, D., Brown, S. & Rand, S. C. Origin of persistent hole burning of N-V centers in diamond. *J. Opt. Soc. Am. B* **9**, 786 (1992).

[13] Manson, N. B. & Wei, C. Transient hole-burning in N-V centre in diamond. *J. Lumin.* **58**, 158 (1994).

[14] Lenef, A. *et al.* Electronic structure of the N-V center in diamond: experiments. *Phys. Rev. B* **53**, 13427 (1996).

[15] Martin, J. P. D. Fine structure of excited $^3E$ state in nitrogen-vacancy centre of diamond. *J. Lumin.* **81** 237 (1999).

[16] Harrison, J., Sellars, M. J. & Manson, N. B. Optical spin polarization of the N-V centre in diamond. *J. Lumin.* **107**, 245 (2004).

[17] Nizovtsev, A. P. *et al.* NV centers in diamond: spin-selective photokinetics, optical ground state spin alignment and hole burning. *Physica B* **340-342,** 106 (2003).

[18] Hanbury Brown, R. & Twiss, R. Q. Correlation between photons in two coherent beams of light. *Nature* **177**, 27 (1956).

[19] Kurtsiefer, C., Mayer, S., Zarda, P. & Weinfurter, H. Stable solid-state source of single photons. *Phys. Rev. Lett.* **85,** 290 (2000).

[20] Beveratos, A., Brouri, R., Poizat, J.-P. & Grangier, P. Bunching and antibunching from single NV color centers in diamond. quant-ph\0010044 (2000).

[21] van Oort, E. & Glasbeek, M. Fluorescence detected level-anticrossing and spin coherence of a localized triplet state in diamond. *Chem. Phys.* **152,** 365 (1991).

[22] Martin, J. P. D. *et al.* Spectral hole burning and Raman heterodyne signals associated with an avoided crossing in the NV centre in diamond. *J. Lumin.* **86,** 355 (2000).

[23] He, X.-F., Manson, N. B. & Fisk, P. T. H. Paramagnetic resonance of photoexcited N-V defects in diamond. I. Level anticrossing in the $^3$A ground state. *Phys. Rev. B* **47,** 8809 (1993).

[24] Smith, W. V., Sorokin, P. P., Gelles, I. L., & Lasher, G. J. Electron spin resonance of nitrogen donors in diamond. *Phys. Rev.* **115**, 1546 (1959).

[25] Holliday, K., Manson, N. B., Glasbeek, M. & van Oort, E. Optical hole-bleaching by level anti-crossing and cross relaxation in the N-V centre in diamond. *J. Phys. C* **1**, 7093 (1989).

[26] van Oort, E. & Glasbeek, M. Cross-relaxation dynamics of optically excited N-V centers in diamond. *Phys. Rev. B* **40**, 6509 (1989).

[27] van Oort, E., Stroomer, P., & Glasbeek, M. Low-field optically detected magnetic resonance of a coupled triplet-doublet defect pair in diamond. *Phys. Rev. B* **42**, 8605 (1990).

[28] Farrer, R. G. On the substitutional nitrogen donor in diamond. *Solid State Comm.* **7**, 685 (1969).

[29] Butler, J. E. *et al*. Exceptionally high voltage Schottky diamond diodes and low boron doping. *Semicond. Sci. Technol.* **18,** S76 (2003).

[30] Meijer, J. *et al.* Generation of single colour centers by focussed nitrogen implantation. cond-mat/0505063.

[31] Kaiser, W. & Bond, W. L. Nitrogen, a major impurity in common type I diamond. *Phys. Rev.* **115,** 857 (1959).

## Figure Legends

Figure 1. **Characterization of single N-V centers. a**, Atomic structure and relevant energy levels of the N-V center. The green "bond" depicts the symmetry axis. **b**, Spatial PL image (20×20 μm) showing emission from multiple N-V centers; NV1 through NV4 (left to right) are marked. The data is taken by raster-scanning the laser across the sample while measuring the average photon count rate $I_{PL}$ at each position. Darker red corresponds to higher $I_{PL}$. **c**, Intensity correlation function $g^{(2)}(\tau)$ versus $\tau$ for NV1, indicating photon emission from a single N-V center. The data is normalized by the detector count rates, the time bin width and total integration time and has not been corrected for background PL. **d**, Optically-detected ESR for NV1: $I_{PL}$ versus microwave frequency $\nu_{MW}$ (circles) with Lorentzian fit (solid line). The laser power is 1 mW for **b**, and 370 μW for **c** and **d**. The samples are at room temperature for all data in this Letter.

Figure 2. **Polarization and magnetic field anisotropy of single N-V centers. a**, Measurement geometry indicating crystal and transition dipole orientations. **b**, Normalized $I_{PL}$ (radial axis) versus laser polarization angle $\phi$ for NV1, NV2 and NV3. Polarization along $[1\bar{1}0]$ corresponds to $\phi$ = 0. A background is subtracted from each curve and data is taken from -90° to 90° and reproduced from 90° to 270°. The dotted lines are guides to the eye. **c**, Upper panel: calculated ground-state spin splitting $\nu$ as a function of B for $\theta = 6^o$ (solid lines) and $54.7^o$ (dashed lines). Middle panel: probability amplitude $|\alpha|^2$ of $|0\rangle_z$ for each spin level at the same two angles. Lower panel: $I_{PL}$ versus B for NV1, NV2, and NV3, with ~ $1^o$ angle between **B** and [111]. The laser power is 370 μW for **b**, and 1 mW for **c**.

Figure 3. **Control of spin level mixing via magnetic field alignment. a**, $I_{PL}$ versus B for NV1 at specified magnetic field angles. **b**, Zoom of data in **a** (points), focusing on the LAC peak with fits (lines) to the model described in the text. Data taken with smaller field steps at $\theta = 0^o$ are included and an angle of $0.2^o$ is used to fit this data to account for residual spin mixing (see text). **c**, $I_{PL}$ versus B for NV4 (points) and fits (lines) at specified field angles. **d**, Normalized amplitude A of the LAC and 500G peaks versus field angle $\theta$ for NV4 and NV1, respectively. The laser power is 1 mW for NV1 and 2.9 mW for NV4.

Figure 4. **Resonant coupling of a single spin to its neighboring spins. a**, $I_{PL}$ versus B for an ensemble of N-V centers at two field angles. Inset: higher resolution field scans around 500 G showing the nitrogen hyperfine structure. **b**, Field scans showing nitrogen coupling to a single N-V center (NV4) at two indicated field angles with 300 μW laser power. **c**, Field scans at two indicated laser powers for NV1 showing broadening of nitrogen peaks with increased power. For **a** (inset), **b** and **c**, a small linear background is subtracted from the data.

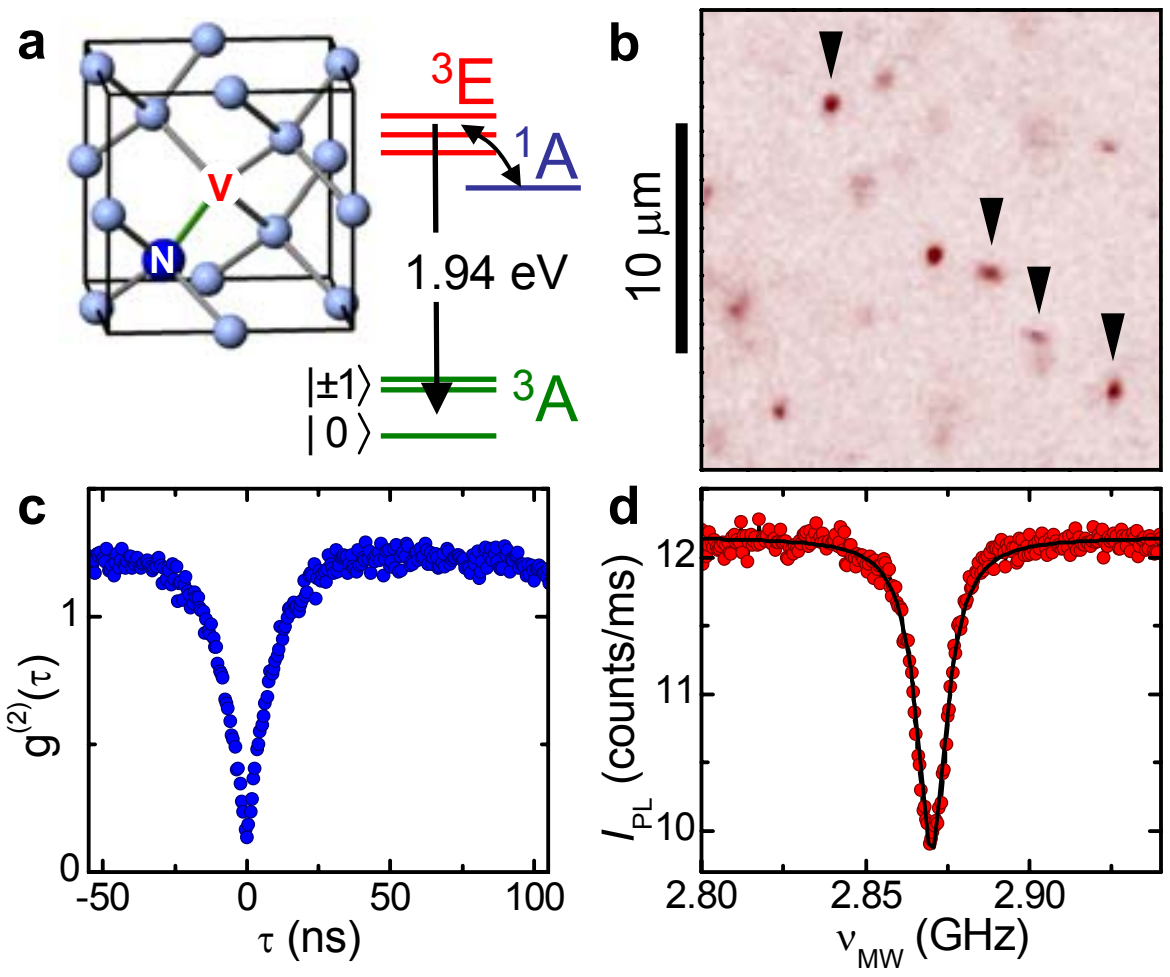

Fig. 1; Epstein et al.

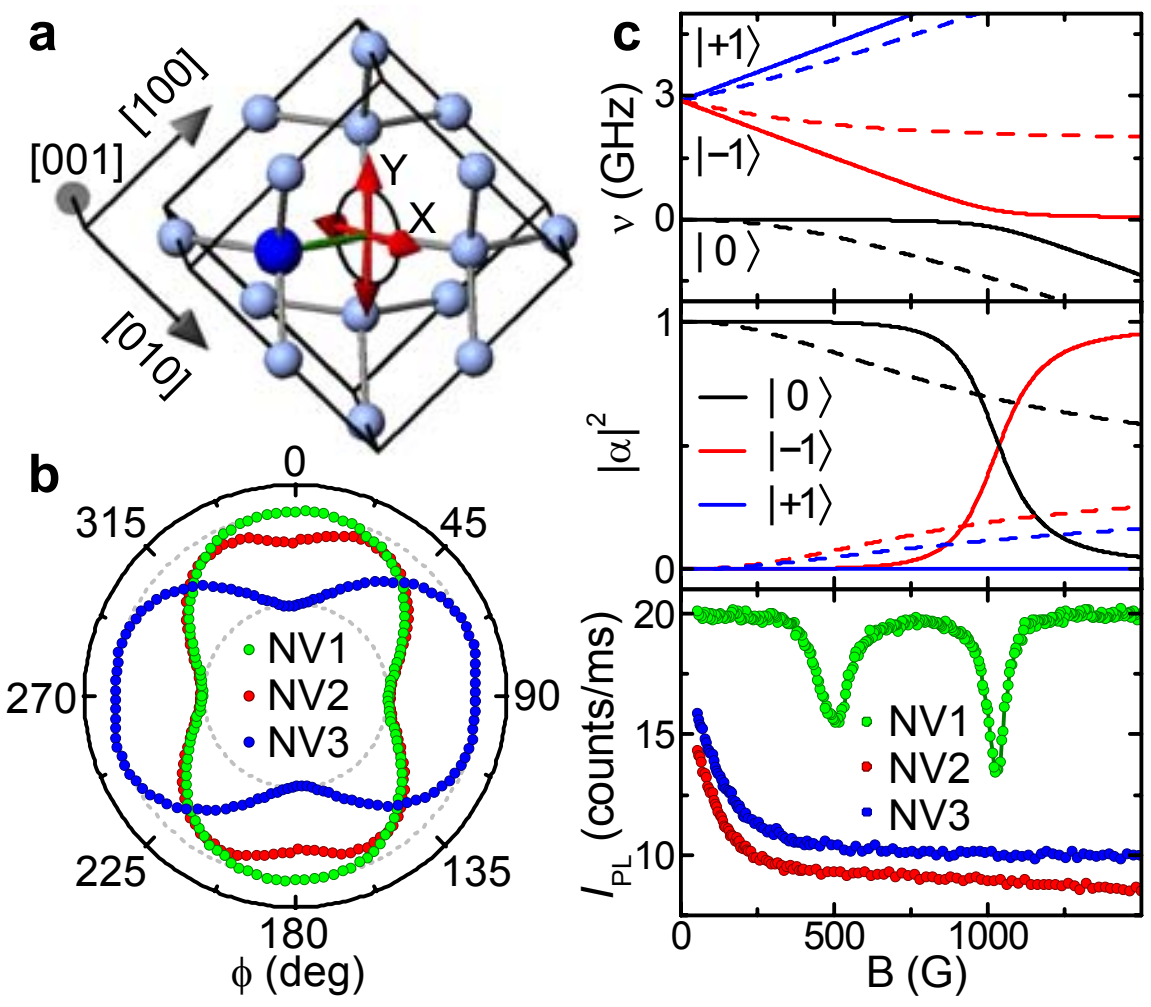

Fig. 2; Epstein et al.

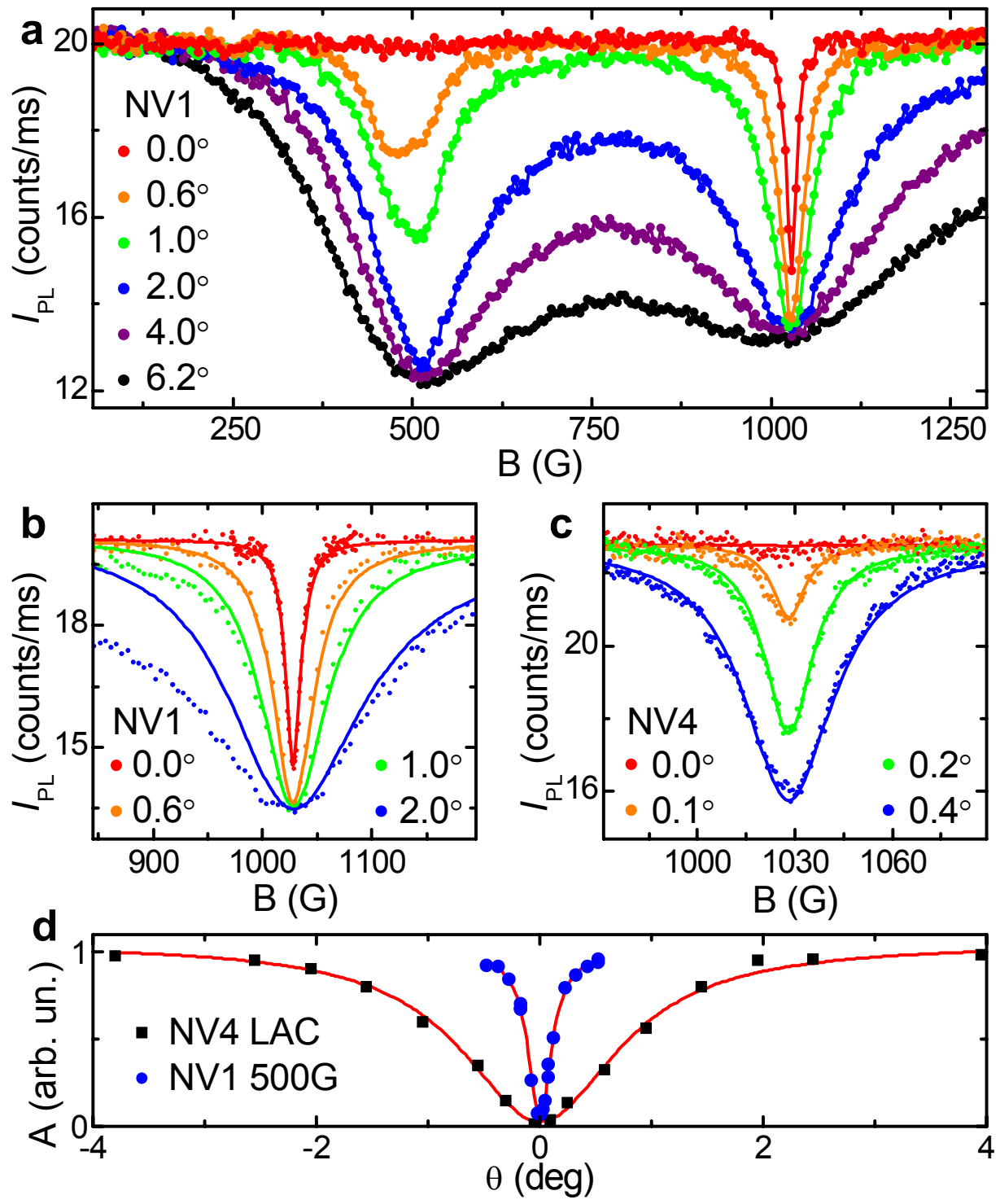

Fig. 3; Epstein et al.

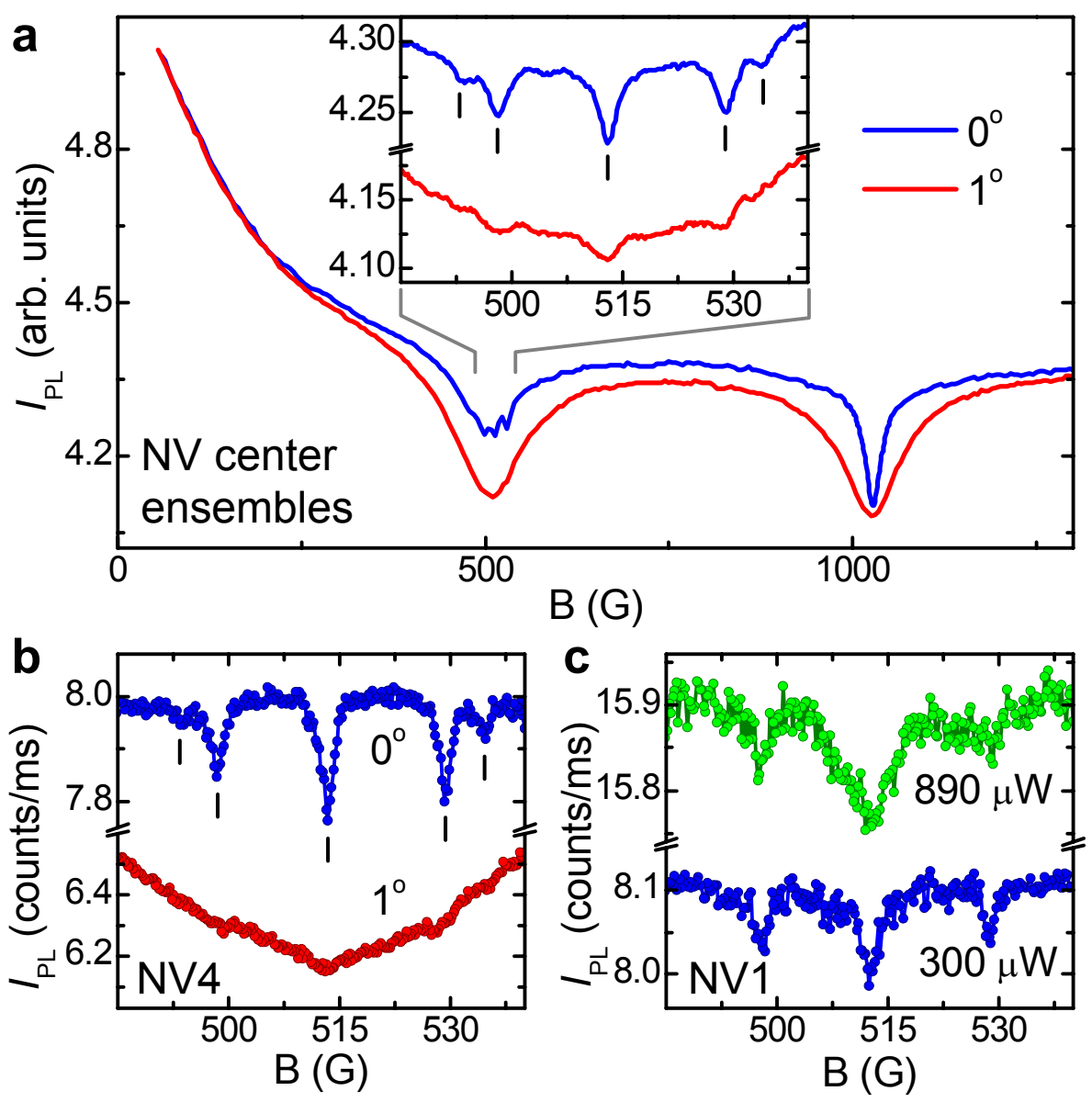

Fig. 4; Epstein et al.